\documentclass[oneside,a4paper]{article}

\usepackage{hyperref}
\usepackage{amsthm}
\usepackage{amssymb}
\usepackage{optidef}

\theoremstyle{definition}
\newtheorem{definition}{Definition}[section]
\newtheorem{theorem}{Theorem}[section]
\theoremstyle{definition}
\newtheorem{exmp}{Example}[section]

\usepackage{authblk}
\providecommand{\keywords}[1]{\textbf{\textit{Keywords:}} #1}

\title{Flexible Skylines: Customizing Skyline Queries Catching Desired Preferences}
\author{Giuseppe Montanaro}
\affil{Politecnico di Milano\\
Milan, Italy\\
\href{mailto:giuseppe.montanaro@mail.polimi.it}{giuseppe.montanaro@mail.polimi.it} }
\date{}

\begin{document}
\maketitle
\begin{abstract}
The techniques most extensively used to retrieve interesting data from data-sets are the Skyline and the Top-k queries. Sadly, they are not enough for facing modern problems, so the needing of something more usable and reliable has come. In this survey we are going to explore Flexible Skylines which are proposed to overcame the old fashion techniques' problems by extending the concept of dominance. After, we are going to compare this approach with the old and new ones evaluating pros and cons. Finally, we will see some interesting applications.
\end{abstract}


\section{Introduction}
The data produced every day is increasing exponentially, especially in last few years, where the use of data is becoming pervasive in almost every application e.g. decision making process, artificial intelligence/machine learning and data mining. At the same time, the size of the produced data is growing as well, so the needing of finding ways to extract relevant data is becoming a crucial aspect for every system. In the classical approach, there are two main path to follow to accomplish this task: \emph{Skyline} and \emph{Top-k queries}. Unfortunately, as argued in \cite{article9, article4} these methods suffer from some drawback. Indeed, the first one is not able to personalize the result and for each user and it feeds the same outcome to everyone. On the other side, with top-k queries, one can actually personalize the output, but this can be done only by a fine tuning process of the weights used by the ranking and this is definitely an hard task to accomplish. In this survey we are going to explore a new technique which tries to merge the benefits of both methods which is called \emph{Flexible Skyline}.


\section{Flexible Skylines}
In this section we are going to explore in detail the the flexible skylines, the concepts behind them and their relevant properties which are extensively discussed in \cite{article1, article2, article13}. First of all, we describe the problem statement and introduce some notation for better dealing with the definitions. 

Consider a relational schema $R$ with the numeric attributes $A_1,...,A_i$, $i > 0$  which are the one we want to consider to carry on the study. For sake of simplicity, we assume that the attributes are already normalized in the interval $[0,1]$. Notice that in this survey, as in the \emph{skyline} related literature, we assume that lower values are considered better than higher ones. A tuple $t$ over $R$ is the function that associates a value $v_i$ with each attribute $A_i$. we will refer to an instance over $R$ as a set of tuples belonging to $R$ with $r$. It should be known to the reader this result stated in \cite{article5} and proofed in \cite{article6}:
$$\forall r,t. t \in sky(r) \Leftrightarrow \exists f \in \mathcal{M}. \forall t' \in r. t \neq t' \Rightarrow f(t) < f(t').$$
Where $\mathcal{M}$ is the set of all monotone scoring functions. Notice that the previous statement is rewritten to fit to our conventions. As stated \cite{article2}, now we can define two operators that extends the concept of \emph{skylines}. The main difference is that the \emph{flexible skyline} operators are applied to subset a $\mathcal{F}$ of $\mathcal{M}$. This lead to the definition of $\mathcal{F}$\emph{-Dominance}.
\begin{definition}[$\mathcal{F}$-Dominance]
Let $\mathcal{F}$ be a subset of monotone scoring functions of $\mathcal{M}$. It is said that a tuple $t$ \emph{F-dominates} another tuple $t'$, if the following condition is satisfied: 
$$\forall f \in \mathcal{F}. f(t) \leq f(t') \wedge \exists f \in \mathcal{F}. f(t) < f(t') .$$
In this case we write $t \prec_F t'$. 
\end{definition}

\begin{exmp}
Consider the tuples $t = (0.1, 0.3, 0.6)$, $s = (0.4, 0.1, 0.5)$, the monotone scoring functions $f_1(x,y)=3x+y$, $f_2(x,y)=x+y+z$ and the set $F =\{f_1,f_2\}$. We have $t \prec_{\mathcal{F}} s$, since $f_2(t) = f_2(s) = 1$ and $f_1(t) = 0.6 < f_1(s) = 1.3$.
\end{exmp}

In particular, we will deal with $\mathcal{F}$ defined as an infinite subset of $\mathcal{M}$ to whom we will later add constrains on the parameters. Additionally, other useful concepts are:
\begin{itemize}
    \item the set of all normalized weighted vectors $\mathcal{W} \subseteq [0,1]^d$. For each vector $W \in \mathcal{W}$, we have that the sum of its components is equal to 1;
    \item the set $\mathcal{C}$ of the linear constraints on the weights;
    \item the set $\mathcal{W}(\mathcal{C}) \subseteq \mathcal{W}$ made out of the elements of $\mathcal{W}$ which satisfy $\mathcal{C}$. 
\end{itemize} 
Moreover, from now on we assume that all the functions in $\mathcal{F}$ are the so called \emph{monotonically transformed, linear-in-the-weights (MLW) function} which can be expressed as follow:
$$f^W(t) = h \left( \sum_{i=1}^d w_i g_i(t[A_i]) \right)$$
Where $W = (w_1, ..., w_d)\in \mathcal{W}(\mathcal{C})$, $t[A_i]$ is the value of tuple \emph{t} on attribute $A_i$, $g_i$ and \emph{h} are either non-decreasing, or non-increasing monotone transforms. We distinguish these two case with the flag $\Lambda$ which is equal 1 in the former case and -1 in the latter.

Now we can move on introducing the first operator: the \emph{non-dominated flexible skyline} which is defined as follow:
\begin{definition}[Non-Dominated Flexible Skyline]
Let $\mathcal{F}$ be a subset of monotone scoring functions of $\mathcal{M}$. The non-dominated flexible skyline on $r$ with respect to $\mathcal{F}$ is the set defined as:
$$ND(r, \mathcal{F}) = \{t \in r. \nexists t' \in r. t' \prec_F t\}$$
\end{definition}
\begin{exmp}
Consider the tuples $t_1 = (2,5), t_2 = (4,4) \text{ and } t_3 = (3,6)$, the instance $r = \{t_1, t_2, t_3\}$, the functions $f_1(x,y)=x+y \text{ and } f_2(x,y)=y$ and the set $\mathcal{F} =\{f_1,f_2\}$. We have that, $ND(r, \mathcal{F})={t_1,t_2}$, since $t_3$ is $\mathcal{F}$-dominated by both $t_1$ and $t_2$.
\end{exmp}
It is reasonable to understand that $\prec_{\mathcal{M}} \equiv \prec$. 

The second operator is called \emph{potentially optimal flexible skylines} which is defined as follow:
\begin{definition}[Potentially Optimal Flexible Skyline]
Let $\mathcal{F}$ be a subset of monotone scoring functions of $\mathcal{M}$. The potentially optimal flexible skyline on $r$ with respect to $\mathcal{F}$ is the set defined as:
$$PO(r, \mathcal{F}) = \{t \in r. \exists f \in \mathcal{F}. \forall t' \in r. t \neq t' \Rightarrow f(t) < f(t')\}$$
\end{definition}
\begin{exmp}
w.r.t. \emph{Example 2.2} the tuple $t_3$ is potentially optimal, since $f_2(t_2) < f_2(t_1)$ and $f_2(t_2) < f_2(t_3)$. Moreover, $t_2$ is potentially optimal as well, since $f_1(t_2) < f_1(t_1)$ and $f_1(t_2) < f_1(t_3)$, thus $PO(r, \mathcal{F}) = \{t_1, t_2\}$.
\end{exmp}
As it could be deduced from the previous examples, the following result can be proofed: 
$$PO \subseteq ND \subseteq sky$$.
This result will be important for the algorithms' implementations as we will see later.

A fundamental concept used to check $\mathcal{F}$-dominance is the $\mathcal{F}$\emph{-dominance region} of a tuple \emph{t} which is defined as follow:
\begin{definition}[$\mathcal{F}$-dominance region]
Let \emph{t} be a tuple and $\mathcal{F}$ be a subset of monotone scoring functions of $\mathcal{M}$. The $\mathcal{F}$\emph{-dominance region} $ND(t, \mathcal{F})$ is the set of point in $[0,1]^d$ that are $\mathcal{F}$-dominated by \emph{t}:
$$ND(t, \mathcal{F}) = \{t' \in [0,1]^d . t \prec_{\mathcal{F}} t'\}.$$
\end{definition}

Now we can explore how the sets of the non-dominated and potentially optimal tuples are computed. Martinenghi and Ciaccia in \cite{article2} state that to find out if $t \prec_{\mathcal{F}} s$ one could solve a linear programming problem in the variables \emph{W} where the objective function is the sum of the differences between the values of the tuples for each attribute transformed by $g_i$ and then weighted by $w_i$. If this LP problem has a positive solution then $t \prec_{\mathcal{F}} s$ otherwise, if the optimal solution is zero, we need to check an additional condition: $t \prec_{\mathcal{F}} s$ if and only if by swapping \emph{s} and \emph{t} we obtain a negative solution. For the other results we have that $t \nprec_{\mathcal{F}} s$. Computing $ND(r, \mathcal{F})$ in such a way could inefficient because the LP problem depends on the pair of tuples' values, if the tuples change the problem is totally different. In fact, another solution is provided and it is based on the concept of $\mathcal{F}$\emph{-dominance region}:
\begin{theorem}[$\mathcal{F}$-Dominance Region]
Let $\mathcal{F}$ be a set of MLW functions subject to a set $\mathcal{C}$ of linear constraints on weights. Let $W^{(1)},..., W^{(q)}$ be the vertices of $\mathcal{W}(\mathcal{C})$. The dominance region of a tuple $\emph{t}$ under $\mathcal{F}$ is the locus of points \emph{s} defined by following the \emph{q} inequalities:
$$\Lambda \cdot \sum_{i=1}^d w_i^{(l)}g_i(s[A_i]) \geq \Lambda \cdot \sum_{i=1}^d w_i^{(l)}g_i(t[A_i]), \quad l \in \{1, ..., q\}$$
And for at least one of them we have:
$$\Lambda \cdot \sum_{i=1}^d w_i^{(m)}g_i(s[A_i]) > \Lambda \cdot \sum_{i=1}^d w_i^{(m)}g_i(t[A_i]), \quad \text{at least one } m \in \{1, ..., q\}$$
\end{theorem}
This method is mush more efficient because the terms of the above expressions are fixed and do not depends on different tuples, so they can just be calculated once.

Speaking about potentially optimal skylines the first important result is the following:
\begin{theorem}[Primal PO Test]
Let $\mathcal{F}$ be a set of MLW functions subject to a set $\mathcal{C}$ of linear constraints on weights. Let $ND(t, \mathcal{F}) = \{t_1, ..., t_{\sigma}, t\}$. Then, $t \in PO(r, \mathcal{F})$ if and only if the following LP problem in the variables $W = (w_1, ..., w_d)$ and $\phi$ has a strictly positive optimal solution:
\begin{maxi*}{}{\phi}{}{}
\addConstraint{}{\Lambda \cdot \textstyle\sum_{i=1}^d w_ig_i(t[A_i]) - w_ig_i(t_j[A_i])+ \phi \leq 0 \quad j \in \{1, ..., \sigma\}}{}
\addConstraint{}{\textstyle\sum_{i=1}^da_{ji}w_i \leq k_j \quad j \in \{1, ..., c\}}{}
\addConstraint{}{w_i \in [0,1] \quad i \in \{1,...,d\}}{}
\addConstraint{}{\textstyle\sum_{i=1}^d w_i = 1}{}
\end{maxi*}
\end{theorem}
Note how, if the LP has a positive optimal solution, then it is sure that a vector $W^*$ exists such that $\textstyle\sum_{i=1}^d w_i^*g_i(t[A_i]) < \textstyle\sum_{i=1}^d w_i^*g_i(t_j[A_i])$ for all the tuples in $ND(t, \mathcal{F})$.

An alternative path could be followed by using the following definition:
\begin{definition}[Convex Combination]
Given tuples $t_1,..., t_n$ a tuples \emph{s} is a \emph{convex combination} of the previous tuples if there exist $\alpha_1,..., \alpha_n$ such that $\alpha_j \in [0,1]$ for $1 \leq j \leq n$, $\sum_{j=1}^n\alpha_j = 1$ and
$$g_i(s[A_i]) = \sum_{j=1}^n \alpha_jg_j(t_j[A_i]), \quad i \in \{1,...,d\}.$$
\end{definition}
This concept leads to the generation of a fictional new tuple who dominates other tuples. The next test uses this notion:
\begin{theorem}[Dual PO Test]
Let $\mathcal{F}$ be a set of MLW functions subject to a set $\mathcal{C}$ of linear constraints on weights. Let $ND(t, \mathcal{F}) = \{t_1, ..., t_{\sigma}, t\}$ and $W^{(1)},..., W^{(q)}$ be the vertices of $\mathcal{W}(\mathcal{C})$. Then, $t \in PO(r;\mathcal{F})$ if and only if there is no convex combinations of $t_1,...,t_{\sigma}$ such that $s\prec_{\mathcal{F}}t$. This happens if the following linear system in the variables $\alpha = (\alpha_1, ..., \alpha_{\sigma})$ is unsatisfiable:
$$\Lambda \cdot \sum_{i=1}^d w_i^{(l)} \left( \sum_{j=1}^{\sigma}\alpha_jg_i(t_j[A_i]) \right) \leq \Lambda \cdot \sum_{i=1}^d w_i^{(l)}g_i(t[A_i])$$
Where $\alpha_j \in [0,1], j \in {1,...,\sigma}$ and $\sum_{j=0}^{\sigma}\alpha_j = 1$
\end{theorem}
This theorem basically states that, if a convex combination (left-hand side of the inequalities) which dominates \emph{t} does not exists, then $t \in PO(r;\mathcal{F})$.

Based on these results, several different algorithms have been proposed. for computing ND we have different options:
\begin{itemize}
    \item \emph{Phases}: in the beginning of the algorithm we can decide if the computation of ND should either starts from \emph{sky} or from scratch;
    \item \emph{Sorting}: we can sort the data-set before the actual algorithm starts on the following rule: \emph{if t precedes s, then s $\nprec_{\mathcal{F}} t$};
    \item $\mathcal{F}$\emph{-Dominance}: for checking if $t \prec_{\mathcal{F}} s$ we can either solve the LP problem or checking if $s \in DR(t,\mathcal{F})$
\end{itemize}
For computing OP we have different options as well:
\begin{itemize}
    \item \emph{Phases}: as before, we can choose either computing $ND(r,\mathcal{F})$ and then subtracting $t \notin OP(r,\mathcal{F})$ or directly eliminate the \emph{non-}$OP(r,\mathcal{F})$ starting from \emph{r};
    \item \emph{PO Test}: we can solve either the primal or the dual problem;
    \item \emph{Incrementality}: since  testing if a tuple is potentially optimal is quite inefficient, we can decide to solve the LP problem with just a subset of the tuples. If the solution indicates that the tuple is not potentially optimal we can eliminate the tuple without problems, otherwise we can just redo the test for an increasing number of tuples.
\end{itemize}
Refer to \cite{article1, article2} for the pseudo-code of the algorithms and for a detailed explanation about the time complexity.

It is important to notice that the framework presented so far could be extended for accepting notions as \emph{top-k} and \emph{k-skyband} with $k > 1$ \cite{article11}.


\section{Comparison}
in \cite{article2} a deep comparison thought different experiments has been developed which shows several important results. Focusing on the normal \emph{skyline queries}, it is shown how the $\mathcal{F}$\emph{-skyline} perform much better in every situation. Indeed the provided data states that the largest reduction in the result set amount to 9.8\% of the normal skyline for ND and to 1\% for PO. Moreover, it seems that the performance improves with increasing data-set dimension, unfortunately a cause cannot be found at the moment. We can also see how the performance improves with increasing number of constraints; unlike the previous case, this result was quite predictable because the constraints affect the set $\mathcal{F}$ and thus reducing the the number of tuples in the result set. Good result have been obtained with different dimensions as well. With $d > 2$ the worst result has still a 35\% value of the ratio \emph{ND/sky} and a remarkable result of 0.38\% for the ration \emph{PO/sky} is archived with $d=10$. Finally, the experiments' results show how the execution time of the normal skyline queries algorithm is still slightly better then the ND and OP ones, but the difference is very little and it does not affect the overall time that much. So the new skyline operators could definitely substitute the normal one without affecting the performance and potentially enhancing the final result.

Unfortunately, a in depth comparison through experiments between top-k queries and the new operators is still missing but some intuitions can be provided considering \emph{precision} and \emph{recall}. In \cite{article2} is illustrated how for all the values of \emph{k} in each dataset we have $pre(sky) \leq pre(ND) \leq pre(OP)$. So the intersection of top-k and skyline will be defintely larger than the ones with NP and OP. As it could be expected, like skyline queries, ND and PO are a lot slower than the top-k queries.


\section{Related works}
In \cite{article7} there is an interesting application of $\mathcal{F}$\emph{-dominance} for the feature selecting problem. Other fascinating fields where the flexible skylines could be applied when the weights of a scoring function are learned via crowdsourcing tasks \cite{article12}. Moreover it could be applied in every fields which are already adopting skyline queries like: economics or stock exchange systems. 

The concept of flexible skylines has been extended by other interesting works. In \cite{article3} the ND and OP operators have been extended as follow:
\begin{itemize}
    \item $ND_k$: it is computed as the set of tuples that are $\mathcal{F}$-dominated by less then k tuples;
    \item $PO_k$: it is computed as the set of tuples which are in the top \emph{k} results for some $f \in \mathcal{F}$.
\end{itemize}
Notice that if $ND_1 = ND$ and $OP_1 = OP$.

Another similar approach which tries to merge the skyline with some kind of user-defined preference, is described in \cite{article10}. It illustrate the \emph{prioritized skyline} (\emph{p-skyline}) that is computed based on the notion of \emph{attribute preference relation}. It define a total order relation on the attribute space through which a priority between the attributes values is established. Starting from that, the concept of \emph{p-expression} is defined as an attribute preference relation. Then, the p-skyline is built by the p-expressions in which every attribute preference is used at most one.

The main downside of the approaches presented so far is that the size of the output cannot be predicted. This could lead to problems when we have to deal with large data-set. To overcame this issue, several options have been proposed. For instance, in \cite{article8} a new concept for dominance is proposed which particularly useful in high dimensional spaces. Indeed, it is stated that \emph{a tuple t k-dominates a tuple s} if there are $k \leq d$ dimensions where \emph{t} is better than or equal \emph{s} and it is better in at least one of these \emph{k} dimensions. In this way they can build the \emph{k-dominant skyline} composed by all the tuples that are not k-dominated by other tuples. Controlling the number of dimensions to be considered can enlarge or dwarf the dimension of the resulting set. 

In \cite{article14}, is described an instance optimal algorithm called \emph{FSA}, which is based on the concept of $\mathcal{F}$\emph{-dominance}, to compute \emph{top-k} queries. This interesting thing about this study is that \emph{FSA} do not require precise values of weights but just estimated ones.

The study which is more related with the ones about flexible skylines is \cite{article4}. In this paper Mouratidis et al. introduce two new operators: \emph{ORD} and \emph{ORU}. These operators have an additional property compared to flexible skylines, which is the possibility to specify the cardinality of the output set. This is obtained using a scoring function (which is basically the inner product between a preference vector \emph{v} of non-negative weights and a tuple) and the concept of $\rho$-\emph{Dominance} defined as follow:
\begin{definition}[$\rho$-\emph{Dominance}]
Let \emph{w} be the best-effort estimate of the user preference vector, called \emph{seed}, and \emph{v} a preference vector of non-negative weights such as $|w -v| \leq \rho$. We say that \emph{a tuple t $\rho$-dominates  a tuple s} if and only if \emph{t} score at least as high as \emph{s} for every \emph{v} and better for at least one of them.
The records that are $\rho$-dominated by fewer than \emph{k} others form the
$\rho$\emph{-skyband}.
\end{definition}
Notice that that larger $\rho$ implies larger $\rho$\emph{-skyband}. In particular, if $\rho = 0$ it collapse to a normal top-k query and if $\rho = \infty$ it collapse to a normal \emph{k-skyband}.

In this situation the operator are defined as follow:
\begin{definition}[ORD]
Let \emph{w} be the seed and \emph{m} the required output size, \emph{ORD} is formed by the records that are $\rho$-dominated by fewer than \emph{k} others, for the minimum $\rho$ that produces exactly \emph{m} records in the output.
\end{definition}

\begin{definition}[ORU]
Let \emph{w} be the seed and \emph{m} the required output size, \emph{ORU} is formed by the records that belong to the top-k result for at least one preference vector within distance $\rho$ from \emph{w}, for the minimum $\rho$ that produces exactly \emph{m} records in the output.
\end{definition}

In \cite{article4} are also provided implementation details and experiments which state that algorithms' performances are acceptable.


\section{Conclusion}
This survey focused on the flexible skylines which allows to customize the the result set unlike the ordinary skyline queries. During the discussion we explored the theory behind them and the implementation. We have understood the strengths of this approach and the drawback as well. We described how related work tried to answer to same questions and we compared the flexible operators with the traditional skyline and top-k queries.

Lastly, we can assert with conviction that Flexible Skylines definitely worth attention thanks to their good performances and it is not to be excluded that further optimization and fixes to current problems could be found.

\bibliographystyle{plain}
\bibliography{refs.bib}

\begin{thebibliography}{10}

\bibitem{article7}
Marcos V.~N. Bedo, Paolo Ciaccia, Davide Martinenghi, and Daniel de~Oliveira.
\newblock A k-skyband approach for feature selection.
\newblock In {\em Proceedings of the 12th International Conference on
  Similarity Search and Applications (SISAP’19)}, pages 160--168, 2019.

\bibitem{article5}
S.~Börzsönyi, D.~Kossmann, and K.~Stocker.
\newblock The skyline operator.
\newblock In {\em IEEE International Conference on Data Engineering (ICDE)},
  pages 421--430, 2001.

\bibitem{article8}
C.~Y. Chan, H.~V. Jagadish, K.~Tan, A.~K.~H. Tung, and Z.~Zhang.
\newblock Finding k- dominant skylines in high dimensional space.
\newblock In {\em SIGMOD Conference}, pages 503--514, 2006.

\bibitem{article6}
Jan Chomicki, Paolo Ciaccia, and Niccolò Meneghetti.
\newblock Skyline queries, front and back.
\newblock {\em SIGMOD Rec. 42, 3 (2013)}, pages 6--18, 2013.

\bibitem{article1}
P.~Ciaccia and D.~Martinenghi.
\newblock Reconciling skyline and ranking queries.
\newblock {\em PVLDB}, 10(11):1454--1465, 2017.

\bibitem{article2}
P.~Ciaccia and D.~Martinenghi.
\newblock Flexible skylines: dominance for arbitrary sets of monotone
  functions.
\newblock {\em ACM Trans. Database Syst.}, 45, 4, Article 18 (December 2020),
  45 pages.

\bibitem{article13}
Paolo Ciaccia and Davide Martinenghi.
\newblock Beyond skyline and ranking queries: Restricted skylines.
\newblock In {\em 26th Italian Symposium on Advanced Database Systems}, pages
  1--8, 2018.

\bibitem{article14}
Paolo Ciaccia and Davide Martinenghi.
\newblock {FA + TA < FSA}: Flexible score aggregation.
\newblock In {\em 27th ACM International Conference on Information and
  Knowledge Management}, pages 57--66, 2018.

\bibitem{article12}
Eleonora Ciceri, Piero Fraternali, Davide Martinenghi, and Marco Tagliasacchi.
\newblock Crowdsourcing for top-k query processing over uncertain data.
\newblock {\em Trans. Knowl. Data Eng}, pages 41--82, 2016.

\bibitem{article9}
A.~A. Freitas.
\newblock A critical review of multi-objective optimization in data mining: a
  position paper.
\newblock {\em SIGKDD Explorations}, 6(2):77--86, 2004.

\bibitem{article10}
D.~Mindolin and J.~Chomicki.
\newblock Preference elicitation in prioritized skyline queries.
\newblock {\em VLDB J.}, 20(2):157--182, 2011.

\bibitem{article4}
Kyriakos Mouratidis, Keming Li, and Bo~Tang.
\newblock Marrying top-k with sky-line queries: relaxing the preference input
  while producing output of controllable size.
\newblock In {\em Proceeding of the 2021 International Conference on Management
  of Data (SIGMOD'21), June 20-25, 2021, Virtual event, China}, 2021.

\bibitem{article3}
Kyriakos Mouratidis and Bo~Tang.
\newblock Exact processing of uncertain top-k queries in multi-criteria
  settings.
\newblock {\em PVLDB}, 11(8):866--879, 2018.

\bibitem{article11}
Dimitris Papadias, Yufei Tao, Greg Fu, and Bernhard Seeger.
\newblock Progressive skyline computation in database systems.
\newblock {\em TODS}, 30(1):41--82, 2005.

\end{thebibliography}
\end{document}